# A simple reform for treating the loss of accuracy of Humlíček's W4 algorithm near the real axis

Mofreh R. Zaghloul

Department of Physics, College of Sciences, United Arab Emirates University, Al-Ain, 15551, UAE.

Tel: +971-3713-6324, Fax: +971-3767-1291, email: m.zaghloul@uaeu.ac.ae

We present a simple reform for treating the reported problem of loss-of-accuracy near the real axis of Humlíček's w4 algorithm, widely used for the calculation of the Faddeyeva or complex probability function. The reformed routine maintains the claimed accuracy of the algorithm over a wide and fine grid that covers all the domain of the real part, $x$, of the complex input variable, $z=x+iy$, and values for the imaginary part in the range $y \in [10^{-30}, 10^{30}]$.

## 1. INTRODUCTION

Because of its applications in many fields of physics such as atmospheric radiative transfer, plasma spectroscopy, nuclear physics, nuclear magnetic resonance, etc., the analysis and evaluation of the complex probability function, commonly known as the Faddeyeva function, has gained a wide interest in the literature [1-25]. The function is defined mathematically as the scaled complementary error function for a complex variable and can, therefore, be expressed as,

$$w(z) = e^{-z^2} \text{erfc}(-iz)$$
$$= e^{(-iz)^2} \text{erfc}(-iz) \qquad (1)$$

where $i = \sqrt{-1}$, $z=x+iy$ is a complex argument, and erfc($z$) is the complementary error function. The function is commonly decomposed into real and imaginary parts as

$$w(x+iy) = V(x, y) + i L(x, y) \qquad (2)$$

where $V$ and $L$ are known as the real and imaginary "Voigt functions," since $V(x,y)$ is the Voigt line profile, used in spectroscopy and radiative transfer, resulting from the convolution of the Gaussian profile (as a result of Doppler broadening) and a Lorentzian profile (result of pressure or collision broadening). The imaginary Voigt function $L$ is also useful for various applications and needs to be taken into consideration as well.

The fact that a closed form expression does not exist for the convolution integral defining the Voigt functions or for the closely related Faddeyeva function led to the development of a wide variety of algorithms to numerically evaluate the functions. These algorithms vary significantly in their accuracy and computational speed and the decision for a certain algorithm will depend on the particular application under consideration.

One of the codes widely employed in the solution of practical problems requiring numerous evaluation of the complex probability function is the Humlíček's w4 code [7]. The code is efficient for some applications tolerating low accuracy computations ($10^{-4}$), however, as shown in the literature (see for example [16,25]) the w4 algorithm does suffer from a loss of its claimed accuracy near the real axis. It is also worth mentioning that Humlíček's w4 algorithm is the basis of several other refinements (see for example [15,16,18,23,24]). In most of these references, Humlíček's w4 algorithm is shown to be remarkably more efficient when the parameter *x* is a vector while *y* is a scalar. Even though, a brief survey and benchmarking tests by Shreier [24] considering practical aspects of Fortran and Python implementations demonstrated that programming language, compiler choice, and implementation details influence computational speed and that there is no unique ranking of algorithms.

In this note we provide a simple reform to this problem where the reformed algorithm maintains the claimed accuracy of the algorithm over the whole domain of interest.

## 2. THE W4 ALGORITHM AND THE LOSS OF ACCURACY REGION

In Humlíček's w4 algorithm, the *x-y* plane is divided into four regions where for each region an approximate expression is used in the form of a rational polynomial. The four regions used in the w4 algorithm are shown in Fig. 1. The borders for these regions are also summarized in the second column of Table 1 below. The w4 algorithm calculates both the real and imaginary parts of the Faddeyeva function concurrently.

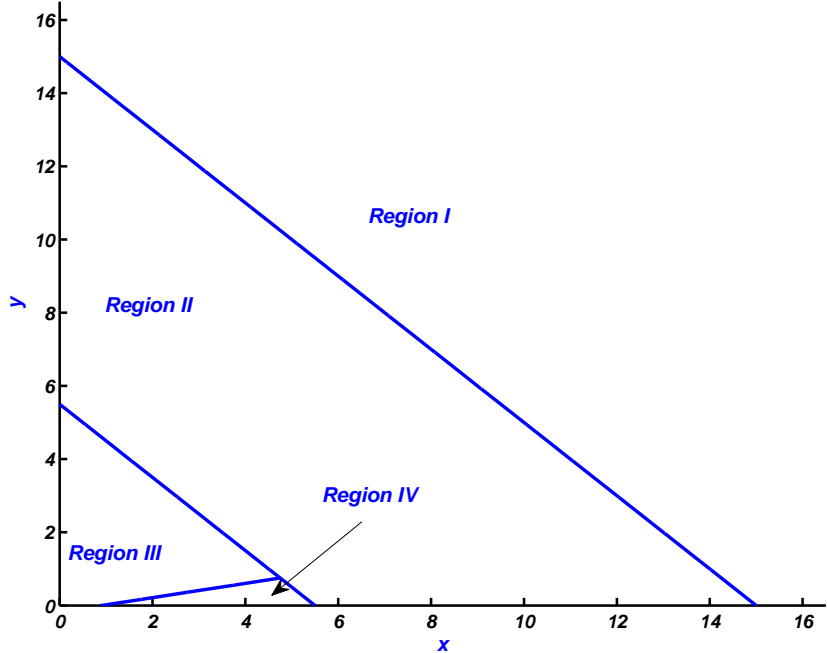

Figure 1: The four regions in the *x-y* plane in the Humlíček's original algorithm

Table 1: Borders of the four regions in Humlíček's original and reformed algorithms

| Region | Original borders of the region | Proposed borders of the region |
|---|---|---|
| I | $\|x\|+y \geq 15.0$ | $\|x\|+y \geq 15.0$ |
| II | $5.5 \leq \|x\|+y < 15.0$ | $5.5 \leq \|x\|+y < 15.0$ **& $y > 10^{-6}$** |
| III | $\|x\|+y < 5.5$ & $y \geq 0.195\|x\|-0.176$ | $\|x\|+y < 5.5$ & $y \geq 0.195\|x\|-0.176$ |
| IV | Otherwise | Otherwise |

A thorough investigation of the accuracy of the w4 code over all of its four regions, using algorithm-916 [25] as a reference, shows that the accuracy failure of the w4 code occurs for values of $y < 10^{-7}$ and values of $x$ in Region II. Figure 2 shows contour plots of the common logarithm (base 10) of the absolute values of the relative error, in the calculation of the real and imaginary parts of the Faddeyeva function, resulting from using the w4 algorithm with algorithm-916 [25] as a reference. As pointed out by Garcia [20] the damping parameter, $y$, for the Lyman transitions of intergalactic $H_I$ spans a range of $9.3 \times 10^{-9}$ (corresponding to higher order Lyman transitions) to $6.5 \times 10^{-4}$ (corresponding to Lyα). Accordingly, this region in which the w4 algorithm loses its claimed accuracy is of practical interest to some applications.

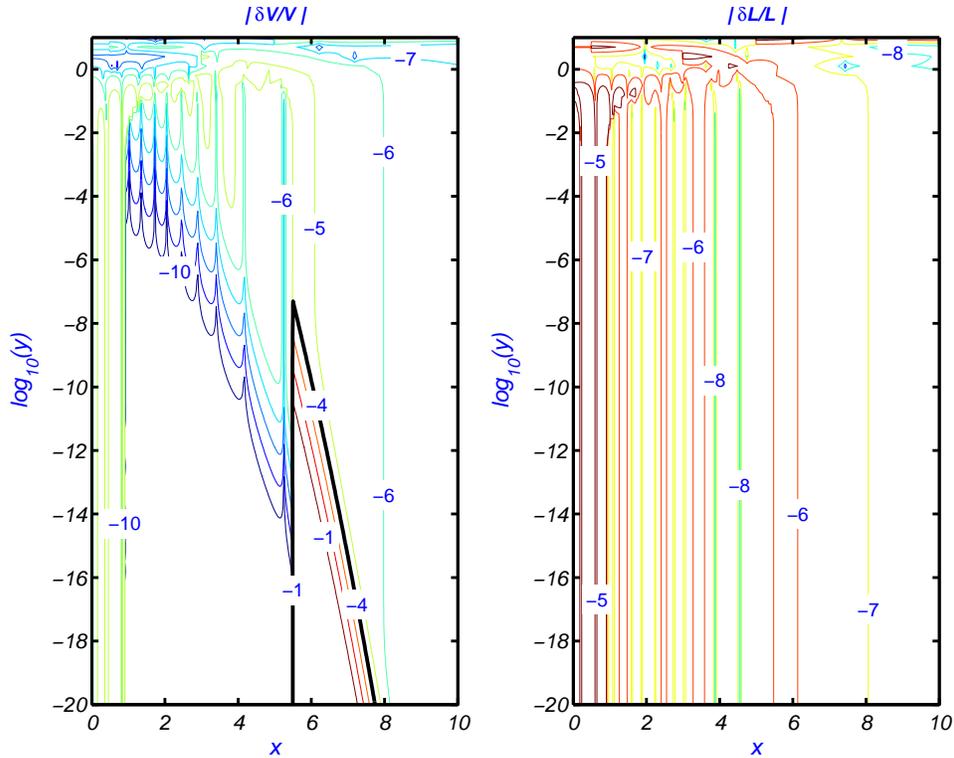

Figure 2: Contour plots of the absolute relative error resulting from using the w4 algorithm for calculating the real and imaginary parts of the Faddeyeva function with algorithm-916 [25] as a reference. The thick line in the left part shows the contour of relative error equals $10^{-4}$ where the region inside this contour is the region where Humlíček's original algorithm loses its claimed accuracy.

## 3. A PROPOSED REFORM

Interestingly a solution to the problem of the loss-of-accuracy of the w4 algorithm near the *x*-axis exists in the reapplication of one of the approximate formulae included in Humlíček's method. The approximation proposed by Humlíček to be used in Region IV in the w4 algorithm is very accurate for very small values of *y*. We propose using this expression for $y \leq 10^{-6}$ and $5.5 \leq |x| + y < 15.0$ to overcome the loss-of-accuracy problem near the real axis. This means that Region IV will be extended to include this part of Region II as schematically shown in Fig. 3. The borders of the modified regions are given in the third column of Table 1 where a simple modification of the borders of Region II is shown in boldface letters. This is a very simple and minor coding change to the original w4 algorithm that can be straightforwardly implemented to fix the above mentioned accuracy problem in such a widely used algorithm.

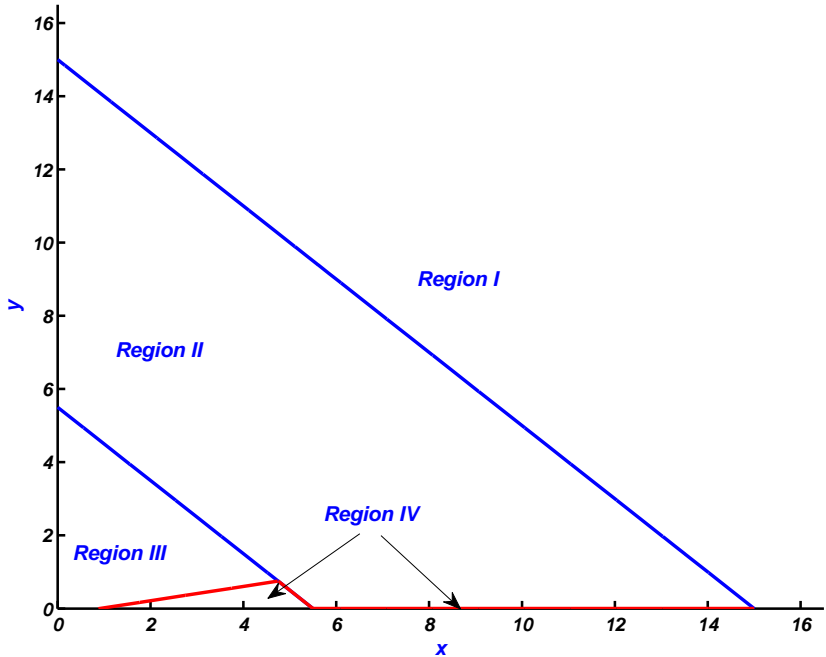

Figure 3: The new four regions in the *x-y* plane in the Humlíček's reformed algorithm

Figure 4 shows contour plots similar to those in Fig. 2 but for results from using the reformed w4 algorithm where, as it can be seen, the claimed accuracy of the w4 algorithm is restored and the loss-of-accuracy problem is removed.

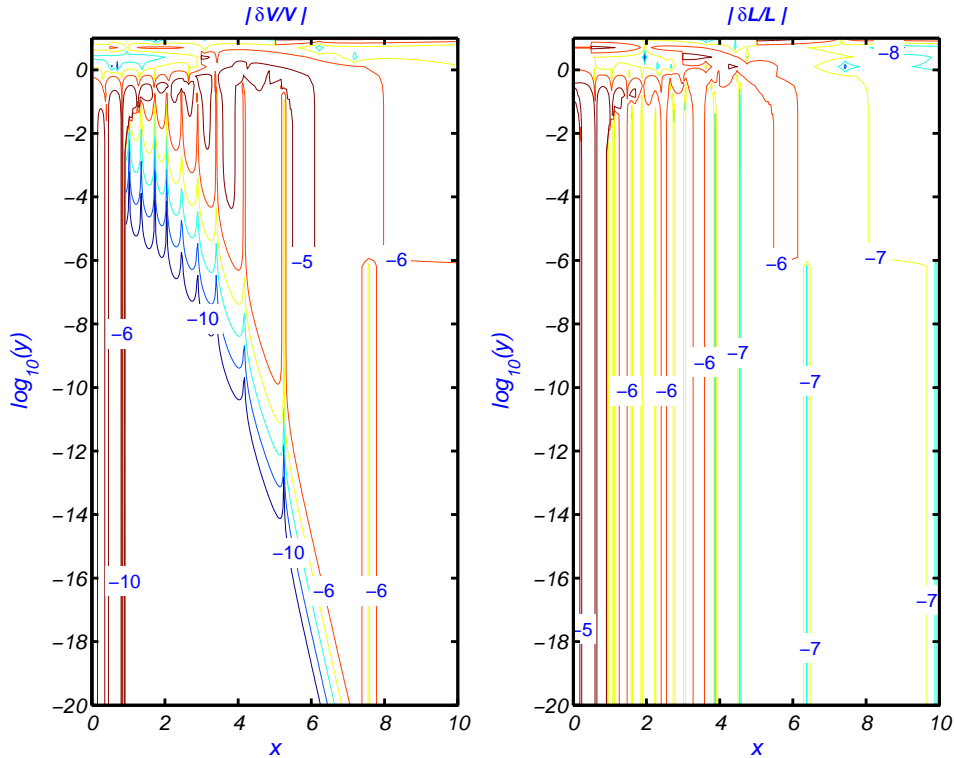

Figure 4: Contour plots of the common logarithm (base 10) of the absolute of the relative error resulting from using the reformed w4 algorithm for calculating the real and imaginary parts of the Faddeyeva function with algorithm-916 [25] as a reference.

## 4. CONCLUSIONS

A simple reform for treating the reported problem of loss-of-accuracy near the real axis of Humlíček's w4 algorithm is introduced. The reformed routine maintains the claimed accuracy of the algorithm over a wide and fine grid that covers all regions of interest.

## ACNOWLEGMENTS

The author would like to acknowledge insightful comments and suggestions received from the reviewers.

## REFERENCES

[1] Faddeyeva, V. N. And Terent'ev, N. M. *Tables of values of the function* $\mathsf{w}(z) = e^{-z^2}\left(1 + \frac{2i}{\sqrt{\pi}} \int_0^z e^{t^2} dt\right)$ *for complex argument*, Gosud. Izdat. The.-Teor. Lit., Moscow, 195; English Transl., Pergamon Press, New York 1961.